\documentclass[reprint,aps,prl,superscriptaddress,nobalancelastpage]{revtex4-2}

\usepackage{graphicx}
\usepackage{dcolumn}
\usepackage{bm}
\usepackage{upgreek}
\usepackage{bm}
\usepackage{amsmath}
\usepackage{color}
\usepackage{multirow}

\usepackage{color}
\usepackage{ulem}

\begin{document}

\title{Viscoelasticity Enhances Contactless Adhesion of Soft Substrates}

\author{Marco Rizzo}
\affiliation{Laboratoire Interfaces \& Fluides Complexes, Universit\'{e} de Mons, 20 Place du Parc, B-7000 Mons, Belgium}

\author{Jacco H. Snoeijer}
\affiliation{Physics of Fluids Group, University of Twente, 7500 AE Enschede, The Netherlands}

\author{Vincent Bertin}
\email[]{Vincent.bertin@univ-cotedazur.fr}
\affiliation{Universit\'{e} C\^{o}te d'Azur, CNRS, INPHYNI UMR 7010, Nice 06200, France}

\author{Pascal Damman}
\email[]{Pascal.damman@umons.ac.be}
\affiliation{Laboratoire Interfaces \& Fluides Complexes, Universit\'{e} de Mons, 20 Place du Parc, B-7000 Mons, Belgium}

\date{\today}

\begin{abstract}
Understanding adhesion is essential for describing stability, friction, and interfacial dynamics. Here, we investigate the adhesion force dynamics between a rigid sphere and a soft surface without direct contact, mediated by a viscous fluid.
By combining controlled experiments, a first-principles visco-elastohydrodynamic theory, and numerical simulations, we demonstrate that viscoelastic relaxation fundamentally modifies elastohydrodynamic adhesion. Rather than simply dissipating energy, viscoelasticity causes the substrate to behave transiently as a stiffer solid, enhancing the maximum adhesive force, changing the early-time force growth for $t^{2/3}$ to $t^{1/3}$, shortening the interaction time, and giving rise to new scaling laws governed by the Deborah number.
The two proposed dimensionless parameters, the softness parameter and the Deborah number, define a unified phase diagram connecting three distinct adhesion regimes: classical Reynolds lubrication, elastohydrodynamic adhesion, and the newly identified visco-elastohydrodynamic regime.

\end{abstract}

\maketitle

\textit{Introduction} --  
Adhesion between solid surfaces is often hindered by surface roughness~\cite{persson2001effect,pastewka2014contact}, which drastically reduces the real area of contact and weakens intermolecular interactions such as van der Waals forces~\cite{israelachvili2011intermolecular}. A classical strategy to overcome this limitation is to use soft materials that deform and conform to surface asperities, thereby increasing the contact area. This mechanism dictates pressure-sensitive adhesives~\cite{creton2003pressure} and is well described by the Johnson–Kendall–Roberts (JKR) theory~\cite{johnson1971surface}, which captures the balance between elasticity and surface energy~\cite{jensen2026physics}. In addition, viscoelastic dissipation in soft materials can significantly enhance adhesion during dynamic detachment, especially at high peeling or separation speeds~\cite{gent1972effect,creton1996does,de1996soft,persson2005crack}.

An alternative route to adhesion relies on the presence of an intervening fluid layer, leading to so-called contactless adhesion~\cite{wang2020dynamic}. This mechanism is widely exploited in biological systems, for instance by insects using adhesive pads~\cite{dirks2011fluid,chen2022bio} or by frogs and chameleons capturing prey with viscous secretions~\cite{brau2016dynamics,noel2017frogs}. In such systems, adhesive forces arise from viscous flows within the confined fluid, as described by classical lubrication theories for rigid surfaces~\cite{stefan1874versuch,reynolds1886iv,bikerman1947fundamentals}. For rigid spherical contacts, however, the predicted adhesive strength depends sensitively on the minimum gap thickness, 
$\delta_0$. The resulting force is therefore not universal and is difficult to predict without an independent knowledge of this small-scale separation. Recent theoretical work has shown that when one of the surfaces is soft, elastic deformation couples to the fluid flow and regularizes the singularity related to vanishing $\delta_0$, giving rise to elastohydrodynamic (EHD) adhesion~\cite{bertin2025sticking}. In the soft limit, this coupling leads to a universal, nonlinear adhesion law 
that shares key features with JKR-like contact, such as the emergence of an apparent contact zone and snapping dynamics, despite the absence of direct solid--solid contact.

Despite these advances, experimental studies of fluid-mediated adhesion for soft substrates remain scarce~\cite{shao2023out}, and theoretical descriptions assume purely elastic materials. However, soft materials are naturally viscoelastic and their response depends on the rate of deformation. This introduces an intrinsic material relaxation time that can compete with the hydrodynamic drainage time~\cite{pandey2016lubrication}. How this competition affects EHD adhesion, and whether it can enhance contactless adhesion in analogy with viscoelastic solid (contact) adhesion, remains an open question.

In this Letter, we investigate EHD adhesion on soft viscoelastic substrates. We show that, while elastic substrates follow existing EHD predictions~\cite{bertin2025sticking}, viscoelasticity leads to an enhancement of adhesion when the substrate relaxation time becomes comparable to the hydrodynamic drainage time. We rationalize this effect with a visco-elastohydrodynamic (V-EHD) model that captures the interplay between lubrication flows and time-dependent elastomer response, revealing a new mechanism of soft viscous adhesion.

\begin{figure*}
    \centering
    \includegraphics[width=\linewidth]{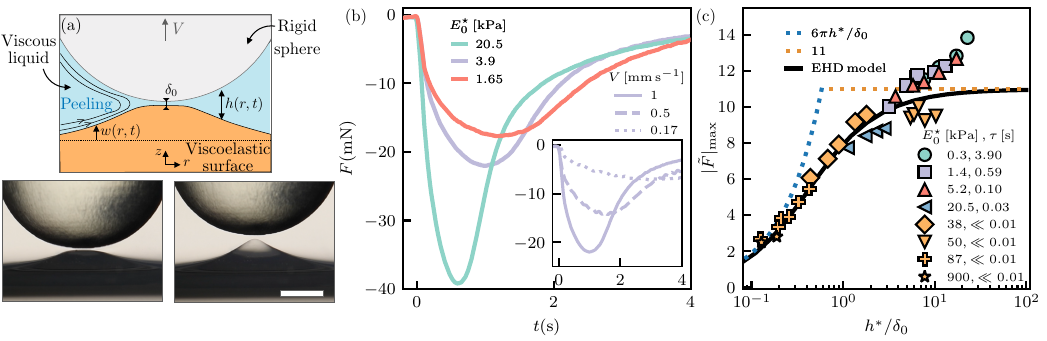}
    \caption{(a) Schematic and experimental images of contactless adhesion with a soft viscoelastic substrate. A rigid sphere is retracted at constant velocity $V$ from an initial gap $\delta_0$ above a viscoelastic substrate. The substrate deformation is denoted $w(r,t)$ and the fluid thickness $h(r,t)$. The images illustrate the loading (left) and peeling (right) stages (Scale bar: 2 mm). (b) Selected curves showing the evolution of the adhesion force with time for different substrate rigidities ($V=1$ mm s$^{-1}$). Inset: Influence of retraction velocities for $E_0^\star=3.9$ kPa. (c) Dimensionless maximum adhesion force as a function of the softness parameter $h^*/\delta_0$. The blue and orange dashed lines correspond to the rigid Reynolds limit and the soft EHD asymptotic limit, respectively, while the solid line is the full numerical EHD prediction.}
    \label{fig:1}
\end{figure*}

\textit{Methods} -- The setup used to study contactless adhesion is shown schematically in Fig.~\ref{fig:1}(a). A rigid sphere of radius $R = 4 \, \mathrm{mm}$ is fully immersed in a viscous liquid of viscosity $62\,\mathrm{Pa\,s}$ and pulled upward at a constant speed $V$ above a solid substrate. The initial separation distance from the bottom surface is denoted $\delta_0$. 
The substrates are layers of cross-linked polydimethylsiloxane, PDMS (Sylgard 184) $2\,\mathrm{cm}$ thick. To tune the viscoelastic properties, we prepare elastomers with different ratios of cross-linking agent. The properties of the PDMS elastomers are measured from dynamic mechanical analysis (DMA)~\cite{jensen2026physics, karpitschka2015droplets} and defined through the reduced elastic modulus $E_0^\star=E_0/(1-\nu^2)$, where $\nu$ and $E_0$ are the Poisson coefficient and the Young modulus together with the characteristic relaxation time, $\tau$. 
The sphere is retracted at a constant velocity in the range $0.17 – 6.7$ mm s$^{-1}$ with initial distance, $20\,\mu\mathrm{m}<\delta_0< 200\,\mu\mathrm{m}$, while the normal force is continuously recorded with a force sensor (Instron). 
Additional experimental details are provided in the Supplemental Materials (SM) S1~\cite{supp}. 

\textit{Force measurements} -- 
Figure~\ref{fig:1}(b) shows typical force versus time curves obtained for different substrate stiffnesses and retraction velocities. In all cases, the force increases to a maximum before decaying. Increasing either the substrate rigidity or the retraction speed leads to higher adhesion forces and shorter times of adhesive interactions. The deformation of the substrate is shown in Fig.~\ref{fig:1}(a) which also illustrates the two stages of the retraction process. During the initial loading phase, or sticking phase, the soft surface follows the sphere displacement, leading to a progressive increase of elastic deformations and adhesive force, up to a maximal surface deformation which coincides roughly with the peak force. Then, in the second phase, the elastic surface relaxes while the adhesive force decreases.
       
For purely elastic substrates, a recent theory predicts that the dynamics is governed by a single dimensionless parameter, $h^*/\delta_0$~\cite{bertin2025sticking}. Here $h^* = R\left({\eta V}/{R E_0^\star}\right)^{2/5}$ is the characteristic elastic deformation resulting from a balance between viscous stresses, $p_\text{vis}\sim \eta R V/h^{*2}$, and elastic stresses, $p_\text{el}\sim E_0^\star h^*/\sqrt{R h^*}$. The corresponding force scale is $F^* = p_\text{vis} Rh^* = \eta R^2 V/h^*$. The ratio $h^*/\delta_0$ determines the effective softness of the substrate for a given measurement. 
We now experimentally check the predictions of the EHD model, focusing on the adhesive force and its maximal value in particular. Two regimes are expected depending on $h^*/\delta_0$.
In the rigid limit, $h^*/\delta_0 \ll 1$, the elastic deformations remain negligible compared to the fluid gap and the response is governed by \textit{rigid} lubrication theory. The pressure then scales as $\eta R V / \delta_0^2$, leading to the classical Reynolds force~\cite{reynolds1886iv}
\begin{equation}
    F = -\frac{6\pi \eta R^2 V}{\delta_0}.
\end{equation}
In the opposite limit $h^*/\delta_0 \gg 1$, elastic deformations exceed the gap thickness [Fig.~\ref{fig:1}(a)], so that the dynamics is more intricate. As we now show, the adhesion becomes independent of $\delta_0$, while the dynamics is then fully controlled by the EHD scales $h^*$ and $F^*$. 

In the initial loading phase, the elastic deformations grow in a self-similar fashion. At the scaling level, the deformation follows the sphere displacement and satisfies $w_\mathrm{st}\sim Vt$ over a radial extent $r_{\mathrm{st}}$, where the subscript st stands for sticking regime. Balancing elastic and viscous pressures, $E_0^\star w_\mathrm{st}/r_\mathrm{st}\sim\eta Vr_\mathrm{st}^2/h_\mathrm{st}^3$, together with the geometrical relation $h_\mathrm{st}\sim r_\mathrm{st}^2/R$, yields $r_\mathrm{st}\sim R[\eta/(E_0^\star t)]^{1/3}$ and therefore
\begin{equation}
\label{eq:early-time_scaling}
F(t \ll t^*)\sim p_\mathrm{st} r_\mathrm{st}^2 \sim VRE_0^{\star 2/3}\eta^{1/3}t^{2/3},
\end{equation}
where $t^*=h^*/V$ is the hydrodynamic drainage time.
This scaling describes the continuous build-up of the adhesive force during the early-time regime. The complete self-similar solution of Ref.~\cite{bertin2025sticking} provides the corresponding numerical prefactor, $\tilde F=-5.94 \,\tilde{t}^{2/3}$, where $\tilde{t}=t/t^*$, and $\tilde{F}=F/F^*$. 

The elastic loading cannot persist indefinitely. As the substrate deformation increases, the assumption that the geometry is governed by the spherical curvature $h_\mathrm{st}\sim r_\mathrm{st}^2/R$ eventually breaks down, once $h_\mathrm{st} \sim w_\mathrm{st}$ or equivalently $r_{st} \sim \sqrt{Rh^*}$. This criterion marks the onset of the second  snapping phase that is characterized by a viscous peeling of the contact edge. This second phase starts at $t \sim t^*$, where we recover the typical EHD time scale $h^*/V$. At the scaling level, the peak force is obtained by inserting this onset time into \eqref{eq:early-time_scaling}, which yields
\begin{equation}
\label{eq:max_force_elastic}
    F_\mathrm{max} \sim F^* = \eta RV \left(\frac{E^\star_0 R}{\eta V}\right)^{2/5}.
\end{equation}
Numerical solutions of the full EHD problem allow the prefactor to be fixed, giving the universal soft-limit value $\tilde F_\mathrm{max}=11$.

\begin{figure}[t]
    \centering
    \includegraphics[width=1.0\columnwidth]{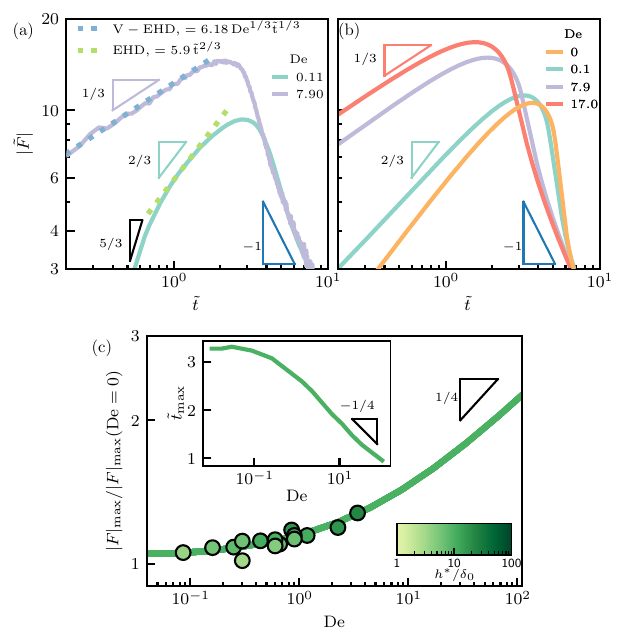}
    \caption{ (a) Dimensionless adhesion force as a function of dimensionless time. The dashed lines show the early-time EHD and V-EHD predictions, Eqs.~\eqref{eq:early-time_scaling} and \eqref{eq:earlytime_VEHD} (b) V-EHD numerical predictions for varying Deborah number. (c) Maximum adhesion force normalized by its elastic value as a function of Deborah number. Symbols are experiments and the solid line is the V-EHD prediction. Inset: dimensionless time at maximum force. In panels (b,c) we set $h^*/\delta_0=10$, corresponding to the soft limit.}
    \label{fig:2}
\end{figure} 

Figure~\ref{fig:1}(c) shows the measured normalized maximum adhesion force, $\tilde F_\text{max}$ as a function of the softness parameter $h^*/\delta_0$. For $E_0^\star > 20$ kPa, the experimental measurements agree with the full range EHD numerical predictions given by the solid black line. The soft asymptotic limit, $\tilde F_\mathrm{max}=11$ is observed for $h^*/\delta_0\gg 1$. The green curve in Fig.~\ref{fig:2}(a) shows that the force curves for $E_0^\star > 20$ kPa follow very nicely the predicted time evolution of the force in $\tilde t^{2/3}$, Eq.~\eqref{eq:early-time_scaling} after a small transient acceleration phase (SM S2 \cite{supp}). This is the first experimental demonstration of the validity of the EHD model proposed in~\cite{bertin2025sticking}.

For very soft substrates ($E_0^\star\leq 5.2$ kPa), however, significant deviations from the EHD model are observed: i) a systematic enhancement of the maximum force [data in Fig.~\ref{fig:1}(c)], ii) a slower increase of the force with time as $\tilde F \sim \tilde t^{1/3}$ [red curve in Fig.~\ref{fig:2}(a)], iii) shorter times of adhesion. 
These deviations suggest that viscoelastic properties of the substrate should be considered. 
Each substrate is characterised by a specific relaxation time $\tau$, ranging from $0.01$ to $3.9$ s (SM S1). We thus introduce the Deborah number that compares the material relaxation time to the characteristic time of the experiment, $\mathrm{De} = \tau/t^* = \tau V / h^*$. Indeed, the deviations from the EHD model arise for Deborah numbers of order unity or larger, confirming that substrate viscoelasticity is the missing physical ingredient. We therefore develop a visco-elastohydrodynamic (V-EHD) model to describe this new regime.

\textit{V-EHD Model} -- The visco-elastohydrodynamic model is based on the following assumptions: (i) small deformations (linear response), (ii) the substrate behaves as a viscoelastic half-space, (iii) a thin fluid gap (lubrication approximation). The local fluid thickness $h$ is therefore
\begin{equation}
h(r,t)=\delta_0+Vt+\frac{r^2}{2R}-w(r,t),
\label{VEHD model Eq 1}
\end{equation}
where $w$ is the substrate deformation [Fig.~\ref{fig:1}(a)].
Within the lubrication approximation, the fluid drainage in the liquid gap obeys the thin-film equation
\begin{equation}
\frac{\partial h(r,t)}{\partial t}=\frac{1}{12\eta r}\frac{\partial}{\partial r}\left(rh^3(r,t)\frac{\partial p(r,t)}{\partial r}\right),
\label{VEHD model Eq 2}
\end{equation}
where $p$ is the hydrodynamic pressure.
The substrate deformation follows from linear viscoelasticity. Using the Boltzmann superposition principle~\cite{lakes2009viscoelastic,karpitschka2015droplets,pandey2016lubrication}, the displacement field reads
\begin{equation}
\label{VEHD model Eq 3}
w(r,t)=-\frac{4}{\pi}
\int_0^\infty \mathcal{M}(r,r')\,dr'
\int_0^t J(t-t')\frac{\mathrm{d}p(r',t')}{\mathrm{d}t'}\,\mathrm{d}t' 
\end{equation}
where $J(t)$ is the creep function and $\mathcal{M}(r,r') = r'/(r+r')K(4rr'/(r+r')^2)$ the axisymmetric elastic Green function of a half-space~\cite{davis1986elastohydrodynamic}, with $K$ the complete elliptic integral of the first kind. 

The rheology of the PDMS obtained from DMA measurements follows a power-law viscoelastic response, captured by a Chasset–Thirion model $E^*(\omega)=E_0^\star (1+(i\omega \tau)^n)$, with $0<n<1$~\cite{rolley2019flexible,zhang2022contactless}. For low crosslinks densities, the exponents are very close to $n = 1/2$ (see SM S1), a value commonly reported~\cite{winter1986analysis,lange1999gelation}. The creep function thus becomes, 
\begin{equation}
\label{eq:relaxation}
\begin{split}
J(t) &=  \int \frac{1}{i\omega E^*(\omega)} 
\frac{\mathrm{d}\omega}{2\pi} =  \frac{1-e^{t/\tau} \left[1 - \text{erf}\left(\sqrt{t/\tau}\right)\right]}{E^*_0}.
\end{split}
\end{equation}
This rheological model is particularly suitable to describe materials exhibiting a broad spectrum of relaxation times leading to a power-law viscoelastic response, as commonly observed in critical gels, polymer networks, and glassy systems \cite{krajina2017dynamic,negi2014viscoelasticity,abidine2013probing,alcoutlabi2003modeling,aime2018power}.

\begin{figure}[t]
    \centering
    \includegraphics[width=0.9\columnwidth]{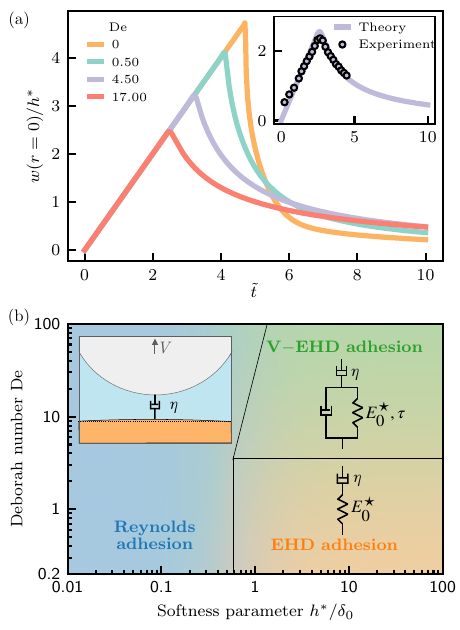}
    \caption{(a) Central substrate deformation as a function of dimensionless time for different Deborah numbers in the soft limit ($h^*/\delta_0\gg1$). Inset: comparison between experiment ($h^*/\delta_0=6.0$, $\mathrm{De}=7.2$) and V-EHD prediction. (b) Phase diagram of contactless adhesion in the $(h^*/\delta_0,\mathrm{De})$ plane. The transition lines are obtained by matching the asymptotic expressions for the maximum adhesion force in the neighboring regimes.}
    \label{fig:3}
\end{figure}

The coupled system described by Eqs.~\eqref{VEHD model Eq 1}-\eqref{VEHD model Eq 3} defines the V-EHD model. We solve this system numerically using a finite-difference scheme for varying Deborah numbers and softness parameter $h^*/\delta_0$ (these are the only dimensionless numbers, see details in SM S3). As shown in Fig.~\ref{fig:2}(b), the V-EHD model captures all experimentally observed trends: the enhancement of the maximum adhesion force, the crossover from the $\tilde{t}^{2/3}$ to $\tilde{t}^{1/3}$ at early times and the decrease of the adhesion time with De.
In addition, the evolution of $\tilde F_\text{max}$ with De is nicely described by the V-EHD model without adjustable parameter [Fig.~\ref{fig:2}(c)]. For a given softness parameter, increasing the viscoelastic strength continuously increases the adhesion force. When $\mathrm{De} \ll 1$, the maximal force closely follows the purely elastic prediction for $F^*$, which depends non-trivially on the softness parameter [black line in Fig.~\ref{fig:1}(c)]. In contrast, when $\mathrm{De} \gg 1$, deviations of the maximal force with respect to the purely elastic model are observed, ultimately reaching an asymptotic scaling $\tilde{F}_{\mathrm{max}} \sim \mathrm{De}^{1/4}$. 
The emergence of this asymptotic scaling can be understood via a simple extension of the previous scaling in the elastic regime. During the beginning of the loading phase, the times are shorter than the relaxation time of the substrate, $t\ll \tau$, and the substrate response can be approximated by a power-law viscoelastic behavior. Therefore, the creep function simplifies to
\begin{equation}
    J\left(\frac{t}{\tau} \ll 1\right) \simeq \frac{2(t/\tau)^{1/2}}{E_0^\star \sqrt{\pi}}.
\end{equation}
At the scaling level, this is equivalent to considering a time-dependent effective elastic modulus $E_\mathrm{eff}(t)\sim E^\star_0 (t/\tau)^{-1/2}$. At long times, the creep function saturates to $J(t/\tau\gg 1)\simeq 1/E_0^\star$, so that the usual relaxed elastic response is recovered. Physically, the fluid therefore probes a transiently stiffer substrate during the early loading phase, $E_\mathrm{eff}\gg E^\star_0$. This transient stiffening enhances the buildup of elastic stresses, leading to a larger adhesion force and shorter interaction time, as observed in Figs.~\ref{fig:2}(b) and \ref{fig:3}.

Replacing the bare elastic modulus in the previous loading phase analysis, i.e. Eq.~\eqref{eq:early-time_scaling}, by the time-dependent effective modulus $E_{\rm eff}(t)$ immediately gives the early-time scaling
\begin{equation}
    \label{eq:earlytime_VEHD}
    \tilde F = -C\,\tilde t^{1/3}\mathrm{De}^{1/3}.
\end{equation}
The complete self-similar analysis, presented in SM S4, confirms that the viscoelastic problem admits the same similarity structure as the purely elastic EHD model. It yields the numerical prefactor $C=5.94\left(\sqrt{\pi}/2\right)^{-1/3} \simeq 6.18$. This prediction is in excellent agreement with the experimental early-time dynamics, as shown by the red dashed lines in Fig.~\ref{fig:2}(a).

The maximum force follows from the same breakdown criterion as in the elastic case: the loading regime ends when the substrate deformation becomes comparable to the local geometrical gap. This gives $\tilde t_{\mathrm{max}}\sim \mathrm{De}^{-1/4}$ and, upon substitution into the early-time force law, $\tilde F_{\mathrm{max}}\sim \mathrm{De}^{1/4}$, in excellent agreement with numerical results of Fig.~\ref{fig:2}(c). The evolution of measured maximum forces with De also quantitatively follow the predictions of the V-EHD model for the entire range of softness parameter [Fig.~\ref{fig:2}(c)]. 

Once the dynamics depart from the self-similar regime, the purely elastic theory predicts a localization of the viscous flow at the edge of the apparent contact, in the form of viscous peeling~\cite{lister2013viscous}, ultimately leading to a violent snapping of the elastic surface~\cite{bertin2025sticking,
berman2019singular}. For purely elastic substrates, De $\ll 1$, this event is observed in the time evolution of the central deformation near $t/t^* = 5$, where the displacement rapidly decays. 
Figure~\ref{fig:3}(a) shows that for large De numbers, substrate viscoelasticity strongly smoothens this snapping and leads to a slower relaxation of the surface deformation due to the delayed stress relaxation of the substrate. The experimental data are convincingly described by the predicted change in snapping dynamics [inset of Fig.~\ref{fig:3}(a)].

\textit{Conclusion} -- We have shown that contactless adhesion arises from the interplay between fluid flow, elastic deformations and substrate relaxation. This competition is governed by two dimensionless numbers $h^*/\delta_0$ (softness) and $\mathrm{De}$ (viscoelasticity), that define the phase diagram shown in Fig.~\ref{fig:3}(b). The resulting phase diagram unifies the three regimes of contactless adhesion: the classical Reynolds regime, the elastohydrodynamic (EHD) regime, and the visco-elastohydrodynamic (V-EHD) regime.

Beyond providing the first experimental validation of the EHD adhesion model, this work establishes visco-elastohydrodynamics as a new regime of contactless adhesion. We anticipate that the present framework will extend beyond contactless adhesion to lubricated interactions involving soft viscoelastic materials, with potential implications for bioadhesion, tactile perception and soft robotic~\cite{peng2021elastohydrodynamic}, as well as lubricated soft matter such as dense suspensions or emulsions.

\textit{Acknowledgements} -- We thank Fabian Brau for the initial adhesion force measurements. M.R. acknowledges financial support from the Fonds de la Recherche Scientifique (FRS--FNRS) through a FRIA PhD fellowship.

\bibliography{biblio}

\end{document}